\documentclass[pra,amsfonts,amssymb,nofootinbib]{revtex4}
\usepackage{amsmath}
\usepackage{amssymb}
\usepackage{graphics}
\usepackage{latexsym}
\usepackage{rotating}
\usepackage{euscript}

\begin{document}

%\input epsf
%%% Blackboard bold "1". Not in the AMS font set.
\def\Bid{{\mathchoice {\rm {1\mskip-4.5mu l}} {\rm
{1\mskip-4.5mu l}} {\rm {1\mskip-3.8mu l}} {\rm {1\mskip-4.3mu l}}}}
%%%

\newcommand{\eL}{{\cal L}}
\newcommand{\half}{\frac{1}{2}}
\newcommand{\J}{\textbf{J}}
\newcommand{\bP}{\textbf{P}}
\newcommand{\G}{\textbf{G}}
\newcommand{\K}{\textbf{K}}
\newcommand{\M}{{\cal M}}
\newcommand{\E}{{\cal E}}
\newcommand{\bu}{\textbf{u}}
\newcommand{\tr}{\mbox{tr}}
\newcommand{\norm}[1]{\left\Vert#1\right\Vert}
\newcommand{\abs}[1]{\left\vert#1\right\vert}
\newcommand{\set}[1]{\left\{#1\right\}}
\newcommand{\ket}[1]{\left\vert#1\right\rangle}
\newcommand{\bra}[1]{\left\langle#1\right\vert}
\newcommand{\ele}[3]{\left\langle#1\left\vert#2\right\vert#3\right\rangle}
\newcommand{\inn}[2]{\left\langle#1\vert#2\right \rangle}
\newcommand{\Real}{\Bid R}
\newcommand{\dmat}[2]{\ket{#1}\!\!\bra{#2}}

\title{Some Applications for an Euler Angle Parameterization of 
$SU(N)$ and $U(N)$}

\author{Todd Tilma}
\affiliation{The Ilya Prigogine Center for Studies in Statistical Mechanics
and Complex Systems \\
Physics Department \\
The University of Texas at Austin \\
Austin, Texas 78712-1081}
\email[Email:]{tilma@physics.utexas.edu}

\author{E.C.G. Sudarshan}
\affiliation{Center for Particle Physics \\
Physics Department \\
The University of Texas at Austin \\
Austin, Texas 78712-1081}
\email[Email:]{sudarshan@physics.utexas.edu}

\date{\today}

\begin{abstract}
Here we apply our $SU(N)$ and $U(N)$
parameterizations \cite{Tilma1,Tilma2,UandCPN} to the question of entanglement in the 
two qubit and qubit/qutrit system.  In
particular, the group operations which entangle a two qubit pure state
will be given, as well as the corresponding manifold that the
operations parameterize.  We also give the volume of this manifold, as
well as the \textit{hypothesized} volume for the set of all entangled
two qubit pure and mixed states.  Extension of this work to the
qubit/qutrit system will also be given.  

\end{abstract}

\maketitle

%\flushleft{03.65.Bz, 03.65.-w, 03.67.L}

\pagebreak

%--------------------------------------------------------------------

%--------------------------------------------------------------------
\section{Introduction}
\label{sec:suCalculationsIntro}

We know that, in general, $N$-dimensional density matrices can be written in
the form
\begin{equation}
\label{nstate}
\rho = \frac{1}{N}\biggl(\Bid_{N} + \sqrt{\frac{N(N-1)}{2}}( \mathbf{n}\cdot \boldsymbol{\lambda})\biggr).\footnote{Note that orthogonal states in this representation are those that have an angle
$
\theta = \cos^{-1}(\frac{-1}{N-1})
$
between them in the corresponding Hilbert space.}
\end{equation}
This representation is a convenient one since for \textit{pure} states it yields,
\begin{equation}
\mathbf{n}\cdot \mathbf{n} = 1, \text{ and } \mathbf{n}\star \mathbf{n} = \mathbf{n},
\end{equation}
where the ``star'' product is defined according to 
\begin{equation}
(\mathbf{a}\star \mathbf{b})_k = \sqrt{\frac{N(N-1)}{2}}\biggl(\frac{1}{N-2}\biggr)d_{ijk} a_i b_j.
\end{equation}
The way to see this is to just use the following relationship from \cite{GellMann,Greiner}
\begin{equation}
\lambda_i \lambda_j = \frac{2}{N}\delta_{ij}\Bid_{N} + (d_{ijk}+i c_{ijk})\lambda_k,
\end{equation}
where the $c_{ijk}$ and the $d_{ijk}$ are the structure coefficients
for the $SU(N)$ Lie algebra in question.
For $N$-dimensional density matrices there exists a simple procedure
to calculate the components of $\mathbf{n}$: beginning with equation
\eqref{nstate} we use the trace condition on the elements of the algebra
\begin{equation}
Tr[\lambda_{i} \cdot \lambda_{j}] = 2 \delta_{ij}
\end{equation}
to generate
\begin{eqnarray}
\label{ni}
n_{i} &=& \sqrt{\frac{1}{2N(N-1)}}* Tr[(N \rho - \Bid_{N})\cdot \lambda_{i}] \nonumber \\
&=& \sqrt{\frac{N}{2(N-1)}}* Tr[\rho\cdot \lambda_{i}]
\end{eqnarray}
We can use this to represent $\mathbf{n}$ in terms of other
parameterizations such as when 
$\rho = U \rho_d U^\dagger$\footnote{Substitution of 
$\rho = U \rho_d U^\dagger$ into equation
\eqref{ni} yields
$$
n_{i} = \sqrt{\frac{N}{2(N-1)}}*Tr[(U \rho_d U^\dagger)\cdot \lambda_{i}]
$$
which upon calculation gives $\mathbf{n}$ as a function of the $N(N-1)$
group parameters (denoted by $\alpha_i$) and the $N-1$ ``rotations''
(denoted by $\theta_i$) parameterizing $\rho_d$.} or to
evaluate \eqref{nstate} when $\rho$ is explicitly given as in the
following example.
%---
%---
\subsection{Example Calculation: Bell's States}

We begin by defining the four Bell states \cite{JSBell1}:
\begin{equation}
\psi_1 = \frac{1}{\sqrt{2}}\begin{pmatrix}
1 \\
0 \\
0 \\
1
\end{pmatrix},
\quad
\psi_2 = \frac{1}{\sqrt{2}}\begin{pmatrix}
1 \\
0 \\
0 \\
-1
\end{pmatrix},
\quad
\psi_3 = \frac{1}{\sqrt{2}}\begin{pmatrix}
0 \\
1 \\
1 \\
0
\end{pmatrix},
\quad
\psi_4 = \frac{1}{\sqrt{2}}\begin{pmatrix}
0 \\
1 \\
-1 \\
0
\end{pmatrix},
\end{equation}
which can also be represented as the following density matrices
\begin{equation}
\label{bstates}
\begin{array}{crcr}
\psi_1\psi_1^\dagger \equiv \rho_{BS1} = \frac{1}{2}\left( \begin{array}{cccc}
                     1 & 0 & 0 & 1 \\
                     0 & 0 & 0 & 0 \\
                     0 & 0 & 0 & 0 \\
                     1 & 0 & 0 & 1  \end{array} \right), &
\psi_2\psi_2^\dagger \equiv \rho_{BS2} = \frac{1}{2}\left( \begin{array}{crcr} 
                     1 &  0 & 0 & -1 \\
                     0 &  0 & 0 & 0 \\
                     0 &  0 & 0 & 0 \\
                     -1 &  0 & 0 & 1  \end{array} \right), \\
\psi_3\psi_3^\dagger \equiv \rho_{BS3} = \frac{1}{2}\left( \begin{array}{crcr} 
                     0 &  0 & 0 & 0 \\
                     0 & 1 & 1 & 0 \\
                     0 & 1 & 1 & 0 \\
                     0 & 0 & 0 & 0  \end{array} \right), &
\psi_4\psi_4^\dagger \equiv \rho_{BS4} = \frac{1}{2}\left( \begin{array}{clcr} 
                     0 & 0 & 0 & 0 \\
                     0 & 1 & -1 & 0 \\
                     0 & -1 & 1 & 0 \\
                     0 & 0 & 0 & 0  \end{array} \right).\\
\end{array}    
\end{equation}
These four density matrices represent the four possible EPR pairs for two qubit systems
in $SU(4)$ and represent an orthonormal basis for the entire two
qubit, pure state, state space (i.\ e.\ for the vector space but not for $\rho$).  By their definition, they
are \textit{maximally} entangled states, i.e. nonfactorizable
superpositions of product states, and thus 
impart non-local correlations between the behavior of the two qubits
that make up these four states.  For, if these states were factorizable
\begin{equation}
\rho = \rho_A \otimes \rho_B,
\end{equation}
then the probability of joint detection would also factorize
\begin{equation}
|\rho|^2 = |\rho_A|^2|\rho_B|^2,
\end{equation}
and thus the measurements would be independent of each other. 

Now although the Bell states cannot be decomposed into a set of
product states, we can decompose their density
matrix representation given in equation \eqref{bstates} into the one
given by equation \eqref{nstate} by 
using equation \eqref{ni} with $N=4$,
\begin{equation}
\rho = \frac{1}{4}(\Bid_{4} + \sqrt{6} \; \mathbf{n}\cdot \boldsymbol{\lambda}),
\end{equation}
thus yielding
\begin{eqnarray}
\rho_{BS1}&=&\frac{1}{4}(\Bid_{4}+\sqrt{6}(\frac{1}{\sqrt{6}}\lambda_3+\frac{1}{3\sqrt{2}}\lambda_8+
\sqrt{\frac{2}{3}}\lambda_9-\frac{1}{3}\lambda_{15})),\nonumber\\
\rho_{BS2}&=&\frac{1}{4}(\Bid_{4}+\sqrt{6}(\frac{1}{\sqrt{6}}\lambda_3+\frac{1}{3\sqrt{2}}\lambda_8-
\sqrt{\frac{2}{3}}\lambda_9-\frac{1}{3}\lambda_{15})),\nonumber\\
\rho_{BS3}&=&\frac{1}{4}(\Bid_{4}+\sqrt{6}(-\frac{1}{\sqrt{6}}\lambda_3+
\sqrt{\frac{2}{3}}\lambda_6-\frac{1}{3\sqrt{2}}\lambda_8+\frac{1}{3}\lambda_{15})),\nonumber\\
\rho_{BS4}&=&\frac{1}{4}(\Bid_{4}+\sqrt{6}(-\frac{1}{\sqrt{6}}\lambda_3-
\sqrt{\frac{2}{3}}\lambda_6-\frac{1}{3\sqrt{2}}\lambda_8+\frac{1}{3}\lambda_{15})).
\end{eqnarray}
We can see that for the first two Bell states the only non-zero 
components of $\mathbf{n}$ are $n_{3}$,
$n_{8}$, $n_{9}$ and $n_{15}$ whereas for the last two Bell states 
one switches $n_{9}$ for $n_{6}$ (with
its corresponding Lie algebra component).  The rest of the components 
of $\mathbf{n}$ are zero.  

%--------------------------------------------------------------------

%--------------------------------------------------------------------
\section{General Entangling Operations on Two Qubits} 
\label{sec:entanglingops}

From (\cite{WootersZy,ZyczkowskiV1,ZyczkowskiV2} and 
references within) we know that entangled (and entangleable) two qubit density matrices must satisfy\footnote{
Note that this is just a necessary condition for a state to either be already
entangled, or to possibly be entangled under some group operation.  Obviously, pure states satisfy this criterion automatically, but
it is the mixed state situation to which this criterion is more often applied.
Separable two qubit density matrices have $Tr[\rho^2] \leq \frac{1}{3}$.}
\begin{equation}
\label{wooterscond}
Tr[\rho^2] > \frac{1}{3}.
\end{equation}
If we start with equation \eqref{nstate}, assigning $N=4$
\begin{equation}
\rho = \frac{1}{4}\biggl(\Bid_{4} + \sqrt{6}( \mathbf{n}\cdot \boldsymbol{\lambda})\biggr),
\end{equation}
then equation \eqref{wooterscond} yields
\begin{equation}
Tr[\rho^2] = Tr\biggl[ \frac{1}{16}\biggl(\Bid_{4} + \sqrt{6}( \mathbf{n}\cdot \boldsymbol{\lambda})\biggr)^2 \biggr] = \frac{1}{4}(1 + 3 \norm{\mathbf{n}}^2) > \frac{1}{3}
\end{equation}
implying that $\norm{\mathbf{n}} > 1/3$ for entangled states.
On the other hand if we begin with
$
\rho = U \rho_d U^\dagger
$
where $U \in SU(4)$, we can see that 
\begin{equation}
Tr[\rho^2] = Tr[(U \rho_d U^\dagger)(U \rho_d U^\dagger)] = Tr[U
\rho_d^2 U^\dagger] = Tr[\rho_d^2] > \frac{1}{3}
\end{equation}
where $\rho_d$ is 
\begin{equation}
\label{rhodbs}
\rho_d = \left( \begin{matrix}
\sin^2(\theta_1)\sin^2(\theta_2)\sin^2(\theta_3) & 0 & 0 & 0\\
0 & \cos^2(\theta_1)\sin^2(\theta_2)\sin^2(\theta_3) & 0 & 0\\
0 & 0 & \cos^2(\theta_2)\sin^2(\theta_3) & 0 \\
0 & 0 & 0 & \cos^2(\theta_3) 
\end{matrix} \right)
\end{equation}
%from \eqref{rhod} 
from \cite{Tilma1}
and we have exploited the
knowledge that $ UU^\dagger = \Bid_4$.  
Evaluation of this trace yields the following
demands on the ranges of $\theta_1$, $\theta_2$, and $\theta_3$ (here expressed as functions of $\sin(\theta_i)$
and $\cos(\theta_i)$):
\begin{equation}
\label{rste}
Tr[\rho_d^2] = {\cos ({{\theta }_3})}^4 + \left( {\cos ({{\theta }_2})}^4 + 
     \frac{1}{4}\left( 3 + \cos (4\,{{\theta }_1}) \right) \,{\sin
     ({{\theta }_2})}^4 \right)
    \,{\sin ({{\theta }_3})}^4> \frac{1}{3}. 
\end{equation}
Depending on the ranges of $\theta_i$ one could either have a pure or
mixed state that would satisfy equation \eqref{wooterscond}.
%---
%---
\subsection{Pure State Entanglement}

One can see that for $\theta_i = \pi/2$, ($i=1,2,3$), equation
\eqref{rste} would be satisfied 
and thus we would have a $\rho_d$ which could be entangled. 
Therefore using equation \eqref{rhodbs} with $\theta_i= {\pi}/{2}$,
($i=1,2,3$), we can define a pure state as\footnote{Remember that any 4 by 4 matrix with one element 
along the diagonal equal to unity and the rest equal to zero will be
invariant under a $U(3)$ subgroup of $SU(4)$ and would therefore
represent a pure state.  The representation given in equation
\eqref{explicitpstate} however is ideally suited
for the following calculations based on the group parameters remaining after
evaluating the coset $SU(4)/U(3)$ (see 
%Chapter \ref{chap:UandCPN} 
\cite{UandCPN}
for more details) since it is those remaining 6 parameters (equally split
between three $\lambda_3$ phases and three $\lambda_{(i-1)^2+1}$
($i=2,3,4$) rotations) that are explicitly contained in the pure state volume
measure given in 
%equation (\ref{eq:dvpurestatesforcp3}) 
\cite{UandCPN}
which parameterizes
the space $\mathbb{C}\mbox{P}^3$ via the corresponding Fubini-Study
metric (see \cite{MByrd2,MByrdp1} for the $SU(3)$ case) which directly acts upon the 
$\rho_{\{1,1\}}$ element.  
This pure
state representation is also consistent with the generalized $\rho_d$,
used in 
%Chapters \ref{chap:su4} and \ref{chap:suN},
\cite{Tilma1,Tilma2}
given previously.}
\begin{equation}
\label{explicitpstate}
\rho_d=
\begin{pmatrix}
1 & 0 & 0 & 0\\
0 & 0 & 0 & 0\\
0 & 0 & 0 & 0\\
0 & 0 & 0 & 0
\end{pmatrix}.
\end{equation}
Using equation \eqref{ni}
we can then
calculate the components of $\mathbf{n}$ in terms of the
twelve $\alpha$ and three $\theta$ parameters when $\rho$ is given by
$\rho = U\rho_d U^\dagger$
and $U \in SU(4)$.  From this it would be possible to 
determine the actual parameter values that
would generate the Bell states
given in equation \eqref{bstates} by solving the fifteen simultaneous 
equations implied by the representation of $\mathbf{n}$ in terms of 
the Euler parameters.  Another, more 
instructive way is to apply successive unitary
operations, $U \in SU(4)$, to the two qubit 
pure state given in equation \eqref{explicitpstate} until one achieves the requisite
Bell state.\footnote{Because of the $U(3)$ invariance of the two qubit pure state given in equation 
\eqref{explicitpstate}, we only have to look at those operations in the coset $SU(4)/U(3)$ which we know
from 
\cite{UandCPN}
%Chapter \ref{chap:UandCPN}
to be represented by 
$$
e^{i\lambda_3 \alpha_1}e^{i\lambda_2 \alpha_2}e^{i\lambda_3 \alpha_3}e^{i\lambda_5 \alpha_4}e^{i\lambda_3 \alpha_5}e^{i\lambda_{10} \alpha_6}
$$
but it is instructive to see, through the use of the full group $SU(4)$, that only those operations generated from the set
$\{\lambda_2, \lambda_5, \lambda_{10}\}$ yield pure state entanglement.} 
%---
\subsubsection{Bell States One and Two}

To begin, we first act upon our pure state with the group operations,
\begin{equation}
\label{case1lambda10}
U =\; e^{i\lambda_{10}\alpha} \text{ and }
U^\dagger =\; e^{-i\lambda_{10}\alpha},
\end{equation}
yielding
\begin{equation}
\label{case1UUdagger}
\rho =\; U\rho_d U^\dagger
=\; e^{i\lambda_{10}\alpha} \rho_d e^{-i\lambda_{10}\alpha},
\end{equation}
which in matrix notation is
\begin{align}
\rho=& \begin{pmatrix}
\cos(\alpha) & 0 & 0 & \sin(\alpha)\\
0 & 1 & 0 & 0\\
0 & 0 & 1 & 0\\
-\sin(\alpha) & 0 & 0 & \cos(\alpha)
\end{pmatrix}\cdot \begin{pmatrix}
1 & 0 & 0 & 0\\
0 & 0 & 0 & 0\\
0 & 0 & 0 & 0\\
0 & 0 & 0 & 0
\end{pmatrix}\cdot \begin{pmatrix}
\cos(\alpha) & 0 & 0 & -\sin(\alpha)\\
0 & 1 & 0 & 0\\
0 & 0 & 1 & 0\\
\sin(\alpha) & 0 & 0 & \cos(\alpha)
\end{pmatrix} \nonumber \\
=& 
\begin{pmatrix}
\cos(\alpha)^2 & 0 & 0 & -\cos(\alpha)\sin(\alpha)\\
0 & 0 & 0 & 0\\
0 & 0 & 0 & 0\\
-\cos(\alpha)\sin(\alpha) & 0 & 0 & \sin(\alpha)^2
\end{pmatrix}.
\label{eq:FGbstate}
\end{align}
Taking the partial transpose of the above density matrix yields
\begin{equation}
\rho^{pt}=\begin{pmatrix}
\cos(\alpha)^2 & 0 & 0 & 0\\
0 & 0 & -\cos(\alpha)\sin(\alpha) & 0\\
0 & -\cos(\alpha)\sin(\alpha)& 0 & 0\\
0 & 0 & 0 & \sin(\alpha)^2
\end{pmatrix},
\end{equation}
which has an eigenvalue decomposition equal to 
\begin{equation}
\label{eigenlambda10}
\{\chi_1,\chi_2,\chi_3,\chi_4\}=\{\cos(\alpha)^2,\;-\cos(\alpha)\sin(\alpha),\;\cos(\alpha)\sin(\alpha),\;\sin(\alpha)^2\},
\end{equation}
and where the constant term in the characteristic polynomial is\footnote{The constant term is just the zeroth order 
coefficient of $\chi$ in the characteristic equation.}
\begin{equation}
\label{detlambda10}
\text{Det}(\rho^{pt}-\Bid_4*\chi)\rightarrow -\cos(\alpha)^4\sin(\alpha)^4.
\end{equation}
Recalling that $\alpha$ varies from $0$ to ${\pi}/{2}$ we can see that for $0
< \alpha < \frac{\pi}{2}$ we have an entangled density matrix $\rho$. In particular we
can see that if $\alpha={\pi}/{4}$ we generate the second Bell state 
\begin{align}
\rho\biggr(\alpha=\frac{\pi}{4}\biggl)=&
\begin{pmatrix}
\cos(\frac{\pi}{4})^2 & 0 & 0 & -\cos(\frac{\pi}{4})\sin(\frac{\pi}{4})\\
0 & 0 & 0 & 0\\
0 & 0 & 0 & 0\\
-\cos(\frac{\pi}{4})\sin(\frac{\pi}{4}) & 0 & 0 & \sin(\frac{\pi}{4})^2
\end{pmatrix} \nonumber \\
=&\begin{pmatrix}
\frac{1}{2} & 0 & 0 & -\frac{1}{2}\\
0 & 0 & 0 & 0\\
0 & 0 & 0 & 0\\
-\frac{1}{2}& 0 & 0 & \frac{1}{2}
\end{pmatrix}.
\end{align}

Now, in general, if
we use 
the Euler angle parameterization of $SU(4)$ given in \cite{Tilma1}
%equation (\ref{eq:su4eas}) 
and take the most general $U \in SU(4)$ to be given as
\begin{align}
U =&\; U(\alpha_1,0,\alpha_3,0,\alpha_5,\alpha_6,\alpha_7,0,\alpha_9,0,\alpha_{11},0,\alpha_{13},\alpha_{14},\alpha_{15}) \nonumber \\
=&\; e^{i\lambda_3 \alpha_1}e^{i\lambda_3 \alpha_3}e^{i\lambda_3 \alpha_5}e^{i\lambda_{10} \alpha_6}e^{i\lambda_3 \alpha_7}e^{i\lambda_3 \alpha_{9}}e^{i\lambda_3 \alpha_{11}}e^{i\lambda_3 \alpha_{13}}e^{i\lambda_8 \alpha_{14}}e^{i\lambda_{15} \alpha_{15}},
\label{eq:case1Ugen}
\end{align}
we would then have a density matrix equal to
\begin{align}
\rho =&\; U\rho_d U^\dagger \nonumber \\
=&\begin{pmatrix}
\cos(\alpha_6)^2 & 0 & 0 & -e^{i(\alpha_1+\alpha_3+\alpha_5)}\cos(\alpha_6)\sin(\alpha_6)\\
0 & 0 & 0 & 0\\
0 & 0 & 0 & 0\\
-e^{-i(\alpha_1+\alpha_3+\alpha_5)}\cos(\alpha_6)\sin(\alpha_6) & 0 & 0 & \sin(\alpha_6)^2\\
\end{pmatrix},
\end{align}
whose partial transpose is
\begin{equation}
\rho^{pt}=
\begin{pmatrix}
\cos(\alpha_6)^2 & 0 & 0 & 0\\
0 & 0 & -e^{-i(\alpha_1+\alpha_3+\alpha_5)}\cos(\alpha_6)\sin(\alpha_6) & 0\\
0 & -e^{i(\alpha_1+\alpha_3+\alpha_5)}\cos(\alpha_6)\sin(\alpha_6) & 0 & 0\\
0 & 0 & 0 & \sin(\alpha_6)^2\\
\end{pmatrix},
\end{equation}
which yields an eigenvalue decomposition and a constant term in the characteristic polynomial equivalent to 
equations \eqref{eigenlambda10} and \eqref{detlambda10} but that does not
generate the second Bell state when $\alpha_6={\pi}/{4}$
\begin{align}
\rho\biggr(\alpha_6=\frac{\pi}{4}\biggl)=&
\begin{pmatrix}
\cos(\frac{\pi}{4})^2 & 0 & 0 & -e^{i(\alpha_1+\alpha_3+\alpha_5)}\cos(\frac{\pi}{4})\sin(\frac{\pi}{4})\\
0 & 0 & 0 & 0\\
0 & 0 & 0 & 0\\
-e^{-i(\alpha_1+\alpha_3+\alpha_5)}\cos(\frac{\pi}{4})\sin(\frac{\pi}{4}) & 0 & 0 & \sin(\frac{\pi}{4})^2\\
\end{pmatrix} \nonumber \\
=&\begin{pmatrix}
\frac{1}{2} & 0 & 0 & -\frac{1}{2}e^{i(\alpha_1+\alpha_3+\alpha_5)}\\
0 & 0 & 0 & 0\\
0 & 0 & 0 & 0\\
-\frac{1}{2}e^{-i(\alpha_1+\alpha_3+\alpha_5)}& 0 & 0 & \frac{1}{2}\\
\end{pmatrix}.
\end{align}
What we would have, on the other hand, is the following. 
Define $\alpha_1+\alpha_3+\alpha_5 = \beta$, where $0 \leq \beta \leq 5\pi$ since
$\alpha_1$ runs from $0$ to $\pi$ and $\alpha_3$ and $\alpha_5$ each run from $0$ to $2\pi$.\footnote{We are now
using the \textit{covering} ranges for $SU(4)$, defined in 
\cite{Tilma1}.}
%Appendix \ref{app:paramranges}.}  
Then
\begin{align}
e^{\pm i(\alpha_1+\alpha_3+\alpha_5)} =&\; e^{\pm i\beta} \nonumber \\
=& \cos(\beta) \pm i\sin(\beta) \nonumber \\
=& -1 \text{ when } \beta=\pi,3\pi,5\pi \nonumber \\
=& 1 \text{ when } \beta=0,2\pi,4\pi.
\end{align}
Thus when $\alpha_1+\alpha_3+\alpha_5=\pi$, $3\pi$, or $5\pi$ we get the first Bell state and
when $\alpha_1+\alpha_3+\alpha_5=0$, $2\pi$, or $4\pi$ we get the second Bell state.  
Intermediate values of $\alpha_1+\alpha_3+\alpha_5$ can be equated to 
``intermediate'' Bell states; states which have an equivalent density
matrix representation as the first and second Bell state, but which
are \textit{not} equal to any type of \textit{convex sum} of said Bell states.

Now if, instead of equation \eqref{case1lambda10} we were to choose
\begin{equation}
U=e^{i\lambda_2\alpha} \text{ or } U=e^{i\lambda_5\alpha}
\end{equation}
and apply it to $\rho_d$ (as in equation \eqref{case1UUdagger}) 
we would not generate the other two Bell states, or for that matter any
entangled density matrix $\rho$, even with the most general U (as in 
equation (\ref{eq:case1Ugen})).  The question now is, what 
combination of the exponentiation of 
$\lambda_2$ with $\lambda_5$, and/or $\lambda_{10}$ will
entangle the pure state density matrix $\rho_d$ yielding the
other two Bell states.  It is to this question 
we now proceed.
%---
\subsubsection{Bell States Three and Four}

To begin, we first act upon our pure state with the group operations,
\begin{equation}
\label{case2lambda52}
U =\; e^{i\lambda_{5}\mu} e^{i\lambda_{2}\nu} \text{ and }
U^\dagger =\; e^{-i\lambda_{2}\nu}e^{-i\lambda_{5}\mu},
\end{equation}
yielding
\begin{equation}
\label{case2UUdagger}
\rho =\; U\rho_d U^\dagger
=\;  e^{i\lambda_{5}\mu} e^{i\lambda_{2}\nu}\rho_d e^{-i\lambda_{2}\nu}e^{-i\lambda_{5}\mu},
\end{equation}
which in matrix notation is\footnote{Recall that since $\lambda_2$ and $\lambda_5$ do \textit{not} commute, the other 
possible group operation $e^{i\lambda_2 \nu}e^{i\lambda_5 \mu}\rho_d e^{-i\lambda_5 \mu}e^{-i\lambda_2 \nu}$ 
will not generate this matrix.  But, because $[\lambda_2, \lambda_5]= - [\lambda_5, \lambda_2]$, 
the subsequent work \textit{after} this step will be similar for either 
$e^{i\lambda_2 \nu}e^{i\lambda_5 \mu}\rho_d e^{-i\lambda_5 \mu}e^{-i\lambda_2 \nu}$ 
or $e^{i\lambda_{5}\mu}e^{i\lambda_{2}\nu}\rho_d e^{-i\lambda_{2}\nu}e^{-i\lambda_{5}\mu}$.}
\begin{align}
\rho=& \begin{pmatrix}
\cos(\mu)\cos(\nu) & \cos(\mu)\sin(\nu) & \sin(\mu) & 0\\
-\sin(\nu) & \cos(\nu) & 0 & 0\\
-\cos(\nu)\sin(\mu) & -\sin(\mu)\sin(\nu) & \cos(\mu) \\
0 & 0 & 0 & 1
\end{pmatrix}\cdot \begin{pmatrix}
1 & 0 & 0 & 0\\
0 & 0 & 0 & 0\\
0 & 0 & 0 & 0\\
0 & 0 & 0 & 0
\end{pmatrix}\cdot \begin{pmatrix}
\cos(\mu)\cos(\nu) & \sin(\nu) & \cos(\nu)\sin(\mu) & 0\\
-\cos(\mu)\sin(\nu) & \cos(\nu) & -\sin(\mu)\sin(\nu) & 0\\
-\sin(\mu) & 0 & \cos(\mu) \\
0 & 0 & 0 & 1
\end{pmatrix} \nonumber \\
=& 
\begin{pmatrix}
\cos(\mu)^2 \cos(\nu)^2 & -\cos(\mu)\cos(\nu)\sin(\nu) & -\cos(\mu)\cos(\nu)^2\sin(\mu) & 0 \\ 
-\cos(\mu)\cos(\nu)\sin(\nu) & \sin(\nu)^2 & \cos(\nu)\sin(\mu)\sin(\nu) & 0 \\ 
-\cos(\mu)\cos(\nu)^2\sin(\mu) & \cos(\nu)\sin(\mu)\sin(\nu) & \cos(\nu)^2\sin(\mu)^2 & 0 \\
0 & 0 & 0 & 0
\end{pmatrix}.
\end{align}

Now the above density matrix does not look like either of the remaining two Bell states; unless we demand that $\mu ={\pi}/{2}$.  Then
we get
\begin{equation}
\label{SGbstate}
\rho=
\begin{pmatrix}
0 & 0 & 0 & 0 \\ 
0  & \sin(\nu)^2 & \cos(\nu)\sin(\nu) & 0 \\ 
0 & \cos(\nu)\sin(\nu) & \cos(\nu)^2 & 0 \\
0 & 0 & 0 & 0
\end{pmatrix},
\end{equation}
which has the same form as the remaining two Bell states.  As before, taking the partial transpose of the above
density matrix yields
\begin{equation}
\rho^{pt}=\begin{pmatrix}
0 & 0 & 0 & \cos(\nu)\sin(\nu)\\
0 & \sin(\nu)^2 & 0 & 0\\
0 & 0 & \cos(\nu)^2& 0\\
\cos(\nu)\sin(\nu) & 0 & 0 & 0
\end{pmatrix},
\end{equation}
which has an eigenvalue decomposition equal to 
\begin{equation}
\label{eigenlambda52}
\{\chi_1,\chi_2,\chi_3,\chi_4\}=\{\cos(\nu)^2,\;-\cos(\nu)\sin(\nu),\;\cos(\nu)\sin(\nu),\;\sin(\nu)^2\},
\end{equation}
and where the constant term in the characteristic polynomial is
\begin{equation}
\label{detlambda52}
\text{Det}(\rho^{pt}-\Bid_4*\chi)\rightarrow -\cos(\nu)^4\sin(\nu)^4.
\end{equation}
Recalling that $\nu$ varies from $0$ to ${\pi}/{2}$ we can see that for $0
< \nu < \frac{\pi}{2}$ we have an entangled density matrix $\rho$. In particular we
can see that if $\nu={\pi}/{4}$ we generate the third Bell state 
\begin{align}
\rho\biggr(\nu=\frac{\pi}{4}\biggl)=&
\begin{pmatrix}
0 & 0 & 0 & 0 \\ 
0 & \sin(\frac{\pi}{4})^2 & \cos(\frac{\pi}{4})\sin(\frac{\pi}{4}) & 0 \\ 
0 & \cos(\frac{\pi}{4})\sin(\frac{\pi}{4}) & \cos(\frac{\pi}{4})^2 & 0 \\
0 & 0 & 0 & 0
\end{pmatrix}, \nonumber \\
=&\begin{pmatrix}
0 & 0 & 0 & 0\\
0 & \frac{1}{2} & \frac{1}{2} & 0\\
0 & \frac{1}{2} & \frac{1}{2} & 0\\
0 & 0 & 0 & 0
\end{pmatrix}.
\end{align}

Notice, if we had instead used
\begin{equation}
U\rho U^\dagger = e^{i\lambda_2 \psi}e^{i\lambda_5 \phi}\rho_d e^{-i\lambda_5 \phi}e^{-i\lambda_2 \psi}
\end{equation}
as our initial starting point, we would have instead produced the following density matrix
\begin{equation}
\rho = 
\begin{pmatrix}
\cos(\phi)^2 \cos(\psi)^2 & -\cos(\phi)^2\cos(\psi)\sin(\psi) & -\cos(\phi)\cos(\psi)\sin(\phi) & 0 \\ 
-\cos(\phi)^2\cos(\psi)\sin(\psi) & \cos(\phi)^2\sin(\psi)^2   & \cos(\phi)\sin(\phi)\sin(\psi) & 0 \\ 
-\cos(\phi)\cos(\psi)\sin(\phi) & \cos(\phi)\sin(\phi)\sin(\psi) & \sin(\phi)^2& 0 \\
0 & 0 & 0 & 0
\end{pmatrix}.
\end{equation}
One can see that in order for this matrix to be equivalent to \eqref{SGbstate}, one must demand that  
$\psi = {\pi}/{2}$, thus yielding
\begin{equation}
\label{SGbstateequiv}
\rho = 
\begin{pmatrix}
0 & 0 & 0 & 0 \\ 
0 & \cos(\phi)^2   & \cos(\phi)\sin(\phi) & 0 \\ 
0 & \cos(\phi)\sin(\phi) & \sin(\phi)^2& 0 \\
0 & 0 & 0 & 0
\end{pmatrix}.
\end{equation}
For simplicity, we shall generalize this group operation and not the original $e^{i\lambda_5\mu}e^{i\lambda_2 \nu}$ calculation.

Therefore, in general, if
we use 
the Euler angle parameterization of $SU(4)$ given in \cite{Tilma1}
%equation (\ref{eq:su4eas}) 
and take $U \in SU(4)$ to be given as
\begin{align}
U =&\; U(\alpha_1,\frac{\pi}{2},\alpha_3,\alpha_4,\alpha_5,0,\alpha_7,0,\alpha_9,0,\alpha_{11},0,\alpha_{13},\alpha_{14},\alpha_{15}) \nonumber \\
=&\; e^{i\lambda_3 \alpha_1}e^{i\lambda_2 \frac{\pi}{2}} e^{i\lambda_3 \alpha_3}e^{i\lambda_5 \alpha_4}e^{i\lambda_3 \alpha_5}e^{i\lambda_3 \alpha_7}e^{i\lambda_3 \alpha_{9}}e^{i\lambda_3 \alpha_{11}}e^{i\lambda_3 \alpha_{13}}e^{i\lambda_8 \alpha_{14}}e^{i\lambda_{15} \alpha_{15}},
\label{eq:case2Ugen}
\end{align}
we would generate the following density matrix
\begin{align}
\rho =&\; U\rho_d U^\dagger \nonumber \\
=&\begin{pmatrix}
0 & 0 & 0 & 0 \\
0 & \cos(\alpha_4)^2 & e^{-i(\alpha_1-\alpha_3)}\cos(\alpha_4)\sin(\alpha_4) & 0 \\
0 & e^{i(\alpha_1-\alpha_3)}\cos(\alpha_4)\sin(\alpha_4) & \sin(\alpha_4)^2 & 0 \\
0 & 0 & 0 & 0
\end{pmatrix},
\end{align}
whose partial transpose is
\begin{equation}
\rho^{pt} =
\begin{pmatrix}
0 & 0 & 0 & e^{i(\alpha_1-\alpha_3)}\cos(\alpha_4)\sin(\alpha_4) \\
0 & \cos(\alpha_4)^2 & 0 & 0 \\
0 & 0 & \sin(\alpha_4)^2 & 0 \\
e^{-i(\alpha_1-\alpha_3)}\cos(\alpha_4)\sin(\alpha_4) & 0 & 0 & 0
\end{pmatrix},
\end{equation}
which yields an eigenvalue decomposition and a constant term in the characteristic polynomial equivalent to 
equations \eqref{eigenlambda52} and \eqref{detlambda52} but that does not
generate the third Bell state when $\alpha_4={\pi}/{4}$
\begin{align}
\rho\biggr(\alpha_4 = \frac{\pi}{4}\biggl)
=&\begin{pmatrix}
0 & 0 & 0 & 0 \\
0 & \cos(\frac{\pi}{4})^2 & e^{-i(\alpha_1-\alpha_3)}\cos(\frac{\pi}{4})\sin(\frac{\pi}{4}) & 0 \\
0 & e^{i(\alpha_1-\alpha_3)}\cos(\frac{\pi}{4})\sin(\frac{\pi}{4}) & \sin(\frac{\pi}{4})^2 & 0 \\
0 & 0 & 0 & 0
\end{pmatrix}\nonumber \\
=&\begin{pmatrix}
0 & 0 & 0 & 0 \\
0 & \frac{1}{2} & \frac{1}{2}e^{-i(\alpha_1-\alpha_3)} & 0 \\
0 & \frac{1}{2}e^{i(\alpha_1-\alpha_3)} & \frac{1}{2} & 0 \\
0 & 0 & 0 & 0
\end{pmatrix}.
\end{align}
What we would have, as before with the second Bell state, is the following. 
Define $\alpha_1-\alpha_3 = \gamma$, where $-2\pi \leq \gamma \leq 0$ since
$\alpha_1$ and $\alpha_3$ again run from $0$ to $\pi$ and $0$ to $2\pi$ respectively.  Then
\begin{align}
e^{\pm i(\alpha_1-\alpha_3)} =&\; e^{\pm i\gamma} \nonumber \\
=& \cos(-|\gamma|) \pm i\sin(-|\gamma|) \nonumber \\
=& \cos(|\gamma|) \mp i\sin(|\gamma|) \nonumber \\
=& -1 \text{ when } |\gamma|=\pi \nonumber \\
=& 1 \text{ when } |\gamma|=0, 2\pi.
\end{align}
Thus when $|\alpha_1-\alpha_3| =\pi$ we get the fourth Bell state and
when $|\alpha_1-\alpha_3|=0$ or $2\pi$ we get the third Bell state.  
Intermediate values of $|\alpha_1-\alpha_3|$ can be thought of 
as ``intermediate'' Bell states; states which have an equivalent density
matrix representation as the third and fourth Bell state, but which
are \textit{not} equal to any type of \textit{convex sum} of said Bell states.
%---
\subsubsection{General Two Qubit Pure State Entanglement}

The natural extension of the previous work is to look at the case when
we use 
the Euler angle parameterization of $SU(4)$ given in \cite{Tilma1}
%equation (\ref{eq:su4eas}) 
and take $U \in SU(4)$ to be given as\footnote{We only need to look at the first 6 group operations of $U$ because they are the ones that 
``parameterize'' the coset $SU(4)/U(3) = \mathbb{C}\mbox{P}^3$ 
%given in Chapter \ref{chap:UandCPN}.}
\cite{UandCPN}.}
\begin{align}
U =&\; U(\alpha_1,\alpha_2,\alpha_3,\alpha_4,\alpha_5,\alpha_6,\alpha_7,0,\alpha_9,0,\alpha_{11},0,\alpha_{13},\alpha_{14},\alpha_{15}) \nonumber \\
=&\; e^{i\lambda_3 \alpha_1}e^{i\lambda_2 \alpha_2} e^{i\lambda_3 \alpha_3}e^{i\lambda_5 \alpha_4}e^{i\lambda_3 \alpha_5}e^{i\lambda_{10} \alpha_6}e^{i\lambda_3 \alpha_7}e^{i\lambda_3 \alpha_{9}}e^{i\lambda_3 \alpha_{11}}e^{i\lambda_3 \alpha_{13}}e^{i\lambda_8 \alpha_{14}}e^{i\lambda_{15} \alpha_{15}}.
\label{eq:caseGUgen}
\end{align}
We would then generate the following density matrix (see $\rho$ on next page).
\begin{sidewaysfigure}
\begin{equation}
\rho =\; U\rho_d U^\dagger = \nonumber
\end{equation}
\begin{equation}
\biggr(
\begin{smallmatrix}
\cos(\alpha_2)^2\cos(\alpha_4)^2\cos(\alpha_6)^2 & -e^{2i\alpha_1}\cos(\alpha_2)\cos(\alpha_4)^2\cos(\alpha_6)^2\sin(\alpha_2)  
& \frac{-1}{2}e^{i (\alpha_1 + \alpha_3 )}\cos(\alpha_2)\cos(\alpha_6)^2\sin(2\alpha_4)
& \frac{-1}{2}e^{i (\alpha_1 + \alpha_3 + \alpha_5 )}\cos(\alpha_2)\cos(\alpha_4)\sin(2\alpha_6) \\ 
-e^{-2i \alpha_1}\cos(\alpha_2)\cos(\alpha_4)^2\cos(\alpha_6)^2\sin(\alpha_2)
& \cos(\alpha_4)^2\cos(\alpha_6)^2\sin(\alpha_2)^2 & e^{-i( \alpha_1 - \alpha_3 )}\cos(\alpha_4)\cos(\alpha_6)^2\sin(\alpha_2)\sin(\alpha_4) 
& e^{-i ( \alpha_1 - \alpha_3 - \alpha_5 )}\cos(\alpha_4)\cos(\alpha_6)\sin(\alpha_2)\sin(\alpha_6) \\ 
-e^{-i ( \alpha_1 + \alpha_3 )}\cos(\alpha_2)\cos(\alpha_4)\cos(\alpha_6)^2\sin(\alpha_4)
& e^{i ( \alpha_1 - \alpha_3 )}\cos(\alpha_4)\cos(\alpha_6)^2\sin(\alpha_2)\sin(\alpha_4) 
& \cos(\alpha_6)^2\sin(\alpha_4)^2 
& e^{i \alpha_5}\cos(\alpha_6)\sin(\alpha_4)\sin(\alpha_6) \\ 
-e^{-i ( \alpha_1 + \alpha_3 + \alpha_5 )}\cos(\alpha_2)\cos(\alpha_4)\cos(\alpha_6)\sin(\alpha_6)
& e^{i ( \alpha_1 - \alpha_3 - \alpha_5 )}\cos(\alpha_4)\cos(\alpha_6)\sin(\alpha_2)\sin(\alpha_6) 
& e^{-i \alpha_5}\cos(\alpha_6)\sin(\alpha_4)\sin(\alpha_6) & \sin(\alpha_6)^2 
\end{smallmatrix}
\biggl).
\label{eq:genSU4rho}
\end{equation}
\end{sidewaysfigure}
One can see immediately that in order to obtain the general form for 
Bell states three and four, $\alpha_2$ and $\alpha_6$ must be set to ${\pi}/{2}$ and zero respectively.  Similarly, in 
order to obtain the general form for Bell states one and two, 
$\alpha_2$ and $\alpha_4$ must be set to zero.  

Generally, though, the two sets of eigenvalues of the partial transpose of equation (\ref{eq:genSU4rho})
\begin{align}
\Psi_{\pm} &=\pm e^{\frac{-i }{2}\left( 2{{\alpha }_1} + {{\alpha }_5} \right) } \cos ({{\alpha }_4})\cos ({{\alpha }_6})\times \Delta, \nonumber \\
\Phi_{\pm} &=\frac{1}{2} \pm \frac{1}{2}e^{\frac{-i}{2} ( 2{{\alpha }_1} + {{\alpha }_5})}{\sqrt{\left(e^{i (2{{\alpha }_1} +  {{\alpha }_5})} - 4{\cos ({{\alpha }_4})}^2{\cos ({{\alpha }_6})}\right)}}
\times \Delta,
\end{align}
where $\Delta$ is
\begin{align}
\Delta =&\;
{\left(e^{i{{\alpha}_5}}\cos({{\alpha}_6})\sin({{\alpha}_2})\sin({{\alpha}_4})+e^{2i{{\alpha}_1}}\cos
({{\alpha}_2})\sin({{\alpha}_6})\right)}^\frac{1}{2}\nonumber\\
&\times {\left(e^{2i{{\alpha}_1}}\cos({{\alpha}_6})\sin({{\alpha}_2})\sin({{\alpha}_4})+e^{i{{\alpha}_5}}\cos({{\alpha}_2})\sin({{\alpha}_6})\right)}^\frac{1}{2},
\end{align}
indicate that the phase parameter $\alpha_3$ does not contribute.
We claim that one only needs three rotations, and one \textit{overall}
phase in order
to carry out a general entangling operation on $\rho_d$. 
This can be seen if one expands the constant term from the characteristic polynomial for this situation
\begin{align}
\text{Det}(\rho^{pt}-\Bid_4*\epsilon)
\rightarrow &-e^{-2i ( 2\alpha_1 + \alpha_5 )}\cos (\alpha_4)^4 \cos (\alpha_6)^4 \nonumber \\
&\times \left( e^{i \alpha_5}\cos (\alpha_6)\sin (\alpha_2)\sin (\alpha_4) + e^{2i \alpha_1}\cos (\alpha_2)\sin (\alpha_6)\right)^2 \nonumber \\
&\times \left( e^{2i \alpha_1}\cos (\alpha_6)\sin (\alpha_2)\sin (\alpha _4) + e^{i \alpha_5}\cos (\alpha_2)\sin (\alpha_6) \right)^2
\end{align}
and defines $\eta = 2\alpha_1-\alpha_5$ as the cumulative phase
\begin{align}
\epsilon^0 \equiv &
-{\cos({{\alpha}_4})}^4{\cos({{\alpha}_6})}^8{\sin({{\alpha}_2})}^4{\sin({{\alpha}_4})}^4\nonumber\\
&-2{e^{-i\eta}}{\cos({{\alpha}_2}){\cos({{\alpha}_4})}^4{\cos({{\alpha}_6})}^7{\sin({{\alpha}_2})}^3
{\sin({{\alpha}_4})}^3\sin({{\alpha}_6})}\nonumber\\
&-2e^{i\eta}\cos({{\alpha}_2}){\cos({{\alpha}_4})}^4{\cos({{\alpha}_6})}^7{\sin({{\alpha}_2})}^3
{\sin({{\alpha}_4})}^3\sin({{\alpha}_6})\nonumber\\
&-4{\cos({{\alpha}_2})}^2{\cos({{\alpha}_4})}^4{\cos({{\alpha}_6})}^6
{\sin({{\alpha}_2})}^2{\sin({{\alpha}_4})}^2{\sin({{\alpha}_6})}^2\nonumber\\
&-{e^{-2i\eta}}{{\cos({{\alpha}_2})}^2{\cos({{\alpha}_4})}^4{\cos({{\alpha}_6})}^6{\sin({{\alpha}_2})}^2
{\sin({{\alpha}_4})}^2{\sin({{\alpha}_6})}^2}\nonumber\\
&-e^{2i\eta}{\cos({{\alpha}_2})}^2{\cos({{\alpha}_4})}^4{\cos({{\alpha}_6})}^6{\sin({{\alpha}_2})}^2
{\sin({{\alpha}_4})}^2{\sin({{\alpha}_6})}^2\nonumber\\
&-2{e^{-i\eta}}{{\cos({{\alpha}_2})}^3{\cos({{\alpha}_4})}^4{\cos
({{\alpha}_6})}^5\sin({{\alpha}_2})\sin({{\alpha}_4})
{\sin({{\alpha}_6})}^3}\nonumber\\
&-2e^{i\eta}{\cos
({{\alpha}_2})}^3{\cos({{\alpha}_4})}^4{\cos({{\alpha}_6})}^5\sin({{\alpha}_2})\sin({{\alpha}_4}){\sin
({{\alpha}_6})}^3\nonumber\\
&-{\cos({{\alpha}_2})}^4{\cos({{\alpha}_4})}^4{\cos({{\alpha}_6})}^4{\sin({{\alpha}_6})}^4.
\end{align}

From the previous work it is obvious that only the parameters $\alpha_{2i}$
($i=1,2,3$) and $\eta$, the overall phase running from $0$ to $2\pi$,
are needed to parameterize all entangling operations that can be done
on an initial pure state (given in equation \eqref{explicitpstate}).
These three rotations and one phase thus can be used to parameterize
the following manifold
\begin{equation}
\label{psentanglingmanifold}
\EuScript{M}_{\text{E}_{SU(4)}} = \frac{SU(4)}{U(3) \times U(1)_{SU(2)_{\eta}}} = \frac{\mathbb{C}\mbox{P}^3}{U(1)_{SU(2)_{2\alpha_1 - \alpha_5}}},
\end{equation}
the volume of which can be found using the material from
\cite{UandCPN}
%Chapter \ref{chap:UandCPN}
\begin{align}
V_{\EuScript{M}_{\text{E}_{SU(4)}}} = V\biggr(\frac{SU(4)}{U(3) \times U(1)_{SU(2)_{\eta}}}\biggl) =
V\biggr(\frac{\mathbb{C}\mbox{P}^3}{U(1)_{2*\lambda_3}}\biggl) &=
\frac{V_{\mathbb{C}\text{P}^3}}{2*V_{U(1)_{\lambda_3}}} \nonumber \\
&=\frac{\pi^3/6}{2\pi \times 2\pi}
\nonumber \\
&= \frac{\pi}{24}.
\label{eq:psentanglingmanifoldvol}
\end{align}

What we have been able to derive in equation (\ref{eq:psentanglingmanifoldvol}) is the volume of the
manifold of all operations on two qubit pure states which produce
entanglement.  Since these operations act upon a pure state (again, see
equation \eqref{explicitpstate}), which is just a point in
$\mathbb{C}\mbox{P}^3$ and thus of measure zero, one may conclude that,
up to the volume of a measure zero set, the volume of the manifold of
all operations on two qubit pure states which produce entanglement is
equivalent to the volume of the set of all entangled two qubit pure
states
\begin{align}
\{\text{Set of All Entangled Two Qubit Pure States}\} &\equiv \EuScript{M}_{\text{E}_{SU(4)}}
 \nonumber \\
V_{\text{Set of All Entangled Two Qubit Pure States}} &\equiv V_{\EuScript{M}_{\text{E}_{SU(4)}}}
= \frac{\pi}{24}.
\end{align}
This volume is less than the numerically estimated value
calculated by Zyczkowski et.\ al.\ in \cite{ZyczkowskiV1}(eq. 30) and referenced
in \cite{ZyczkowskiV2} for the lower bound of the
volume of entangled two qubit states (1-0.863) by approximately 
6 one-thousandths.  

Unfortunately, whereas our calculation was done
with only pure states in mind, the Zyczkowski et.\ al.\ calculation
was numerically done using a mixed state product measure defined
on $U(N)$ and
randomly chosen density matrices \cite{ZyczkowskiV1,ZyczkowskiV2}
which satisfied the Peres-Horodecki criterion for separability \cite{Peres,Horodeckietal}.
Therefore although a tantalizing conclusion, we must concede that
until a more general calculation is performed using our mixed state
product measure 
(defined in \cite{UandCPN})
%given in Chapter \ref{chap:UandCPN}
we cannot definitively state that we
have calculated the exact volume of the set of all entangled two qubit
pure states.
%---
%---
\subsection{Mixed State Entanglement}

Using common mathematical software, one can see that for $\theta_i \neq \pi/2$, ($i=1,2,3$)
equation \eqref{rste} could still be satisfied \textit{if}
\begin{gather}
\sin ({{\theta }_1}) = {\sqrt{\frac{1}{2} + \frac{{\sqrt{32789757}}}{12482}}} \text{ and } \nonumber \\
   1 > \sin ({{\theta }_2}) > - {\frac{1}{\sqrt{2}}}\,
      {\sqrt{\frac{1 + {\sqrt{{\sin ({{\theta }_1})}^2 - {\sin ({{\theta }_1})}^4}}}
          {1 - {\sin ({{\theta }_1})}^2 + {\sin ({{\theta }_1})}^4}}} \text{ and }
   1 > \sin ({{\theta }_3}) > \frac{{\sqrt{3}}}{2} 
\end{gather} 
or
\begin{gather}
  \sin ({{\theta }_1}) = {\sqrt{\frac{1}{2} + \frac{{\sqrt{\frac{785323439}{3}}}}{37446}}} \text{ and }\nonumber \\
   1 > \sin ({{\theta }_2}) > -{\frac{1}{3\sqrt{2}}}\,
      {\sqrt{\frac{9 + {\sqrt{3}}\,{\sqrt{-1 + 28\,{\sin ({{\theta }_1})}^2 - 28\,{\sin ({{\theta }_1})}^4}}}
          {1 - {\sin ({{\theta }_1})}^2 + {\sin ({{\theta }_1})}^4}}} \text{ and }
   1 > \sin ({{\theta }_3}) > \frac{{\sqrt{3}}}{2}
\end{gather} 
or
\begin{gather}
  \frac{1}{{\sqrt{2}}} < \sin ({{\theta }_1}) < {\sqrt{\frac{1}{2} + \frac{{\sqrt{\frac{785323439}{3}}}}{37446}}} \text{ and } \nonumber \\
   \frac{79}{100} < \sin ({{\theta }_2}) \leq - 
   {\frac{1}{3\sqrt{2}}}\,{\sqrt{\frac{9 + {\sqrt{3}}\,{\sqrt{-1 + 28\,{\sin ({{\theta }_1})}^2 - 
                 28\,{\sin ({{\theta }_1})}^4}}}{1 - {\sin ({{\theta }_1})}^2 + {\sin ({{\theta }_1})}^4}}} 
    \text{ and }\nonumber \\
1 > \sin ({{\theta }_3}) > - 
   {\frac{1}{\sqrt{6}}}\,{\sqrt{\frac{3 + {\sqrt{3}}\,{\sqrt{-1 + 4\,{\sin ({{\theta }_2})}^2 - 
                 4\,{\sin ({{\theta }_2})}^4 + 4\,{\sin ({{\theta }_1})}^2\,{\sin ({{\theta }_2})}^4 - 
                 4\,{\sin ({{\theta }_1})}^4\,{\sin ({{\theta }_2})}^4}}}{1 - {\sin ({{\theta }_2})}^2 + 
            {\sin ({{\theta }_2})}^4 - {\sin ({{\theta }_1})}^2\,{\sin ({{\theta }_2})}^4 + 
            {\sin ({{\theta }_1})}^4\,{\sin ({{\theta }_2})}^4}}} 
\end{gather} 
or
\begin{gather}
  1 > \sin ({{\theta }_1}) > 
    {\sqrt{\frac{1}{2} + \frac{{\sqrt{32789757}}}{12482}}} \text{ and }
   1 > \sin ({{\theta }_2}) > \frac{79}{100} \text{ and }
   1 > \sin ({{\theta }_3}) > \frac{{\sqrt{3}}}{2} 
\end{gather} 
or
\begin{gather}
  {\sqrt{\frac{1}{2} + \frac{{\sqrt{32789757}}}{12482}}} > 
   \sin ({{\theta }_1}) > {\sqrt{\frac{1}{2} + \frac{{\sqrt{\frac{785323439}{3}}}}{37446}}} \text{ and }
   1 > \sin ({{\theta }_2}) > \frac{79}{100} \text{ and }
   1 > \sin ({{\theta }_3}) > \frac{{\sqrt{3}}}{2} 
\end{gather} 
or
\begin{gather}
  \frac{1}{{\sqrt{2}}} < \sin ({{\theta }_1}) < {\sqrt{\frac{1}{2} + \frac{{\sqrt{\frac{785323439}{3}}}}{37446}}} \text{ and }\nonumber \\
   1 > \sin ({{\theta }_2}) > - 
   {\frac{1}{3\sqrt{2}}}\,{\sqrt{\frac{9 + {\sqrt{3}}\,{\sqrt{-1 + 28\,{\sin ({{\theta }_1})}^2 - 
                 28\,{\sin ({{\theta }_1})}^4}}}{1 - {\sin ({{\theta }_1})}^2 + {\sin ({{\theta }_1})}^4}}} 
    \text{ and }
1 > \sin ({{\theta }_3}) > \frac{{\sqrt{3}}}{2}.
\end{gather}
Therefore we could generate a $\rho_d$ which could be entangled but would 
no longer be a pure state.  Unfortunately, the fact that $\rho_d$ is
no longer a pure state also means that we
would have to look at the most general $U \in SU(4)$ acting on $\rho_d$
in order to determine which successive unitary operations $U\rho_d
U^\dagger$ would produce entanglement.  This is a rather lengthy and
complicated calculation and is beyond the scope of this
section.

What we can do though is make an educated guess as to the volume of entangled
two qubit mixed states by noticing that it is a
product of the volume of the 3-dimensional symplex of eigenvalues of
$\rho_d$ (with appropriate ranges) and the volume of the flag manifold
$SU(4)/U(1)_{SU(2)} \times U(1)_{SU(3)} \times
U(1)_{SU(4)}$.\footnote{Explained in detail in 
%Chapter \ref{chap:UandCPN}
\cite{UandCPN}.}  These
volumes can be calculated by using 
%equations 
%(\ref{eq:dmu}) and \eqref{dmuranges} from Chapter \ref{chap:UandCPN}
the mixed state product measure given in \cite{UandCPN}
with
the necessary ranges for the eigenvalues (given above) and the covering
ranges for $SU(4)$ 
(given in \cite{Tilma1}).
%(defined in Appendix \ref{app:paramranges}).  
Using this material we arrive at
\begin{equation}
dV_{E_{ms}}=\alpha_s \Lambda_1^{s-1}\Lambda_2^{s-1}\Lambda_3^{s-1}(1-\sum_{i=1}^3\Lambda_i)^{s-1}d\Lambda_1 \ldots d\Lambda_3 \times 
d\biggr(\frac{SU(4)}{U(1)_{SU(2)} \times U(1)_{SU(3)} \times U(1)_{SU(4)}}\biggl)
d\alpha_{12} \ldots d\alpha_1.
\end{equation}
A general (and rather naive) evaluation of this measure for our situation yields
\begin{equation}
\label{vol2qbms}
V_{E_{ms}}= \alpha_s\, \frac{\left( {{a_L}}^s - {{a_U}}^s \right) \,
    \left( {{b_L}}^s - {{b_U}}^s \right) \,
    \left( {{c_L}}^s - {{c_U}}^s \right) \,
    \left( {{d_L}}^s - {{d_U}}^s \right) }{s^4}\times \frac{\pi^6}{12} = \omega\,\frac{\pi^6}{12},
\end{equation}
where $\{a_U,a_L\}$,$\{b_U,b_L\}$,$\{c_U,c_L\}$ and $\{d_U,d_L\}$ are the squared values of the above maximal and minimal ranges
($\{d_U,d_L\}$ comes from $\Lambda_4 \equiv 1-\sum_{i=1}^3\Lambda_i$).  Since
the symplex measure is \textit{assumed} to be non-zero, and using
the work contained in \cite{UandCPN}
%equation (\ref{eq:2qbomega}) from Chapter \ref{chap:UandCPN}
we can \textit{hypothesize} that $\omega$ has the following bounds (dependent on the value of $s$ and
recalling that $\alpha_s > 0$ and $s >0$)
\begin{equation}
0 <  \left( {{a_L}}^s - {{a_U}}^s \right) \,
    \left( {{b_L}}^s - {{b_U}}^s \right) \,
    \left( {{c_L}}^s - {{c_U}}^s \right) \,
    \left( {{d_L}}^s - {{d_U}}^s \right) < 4^{-4s}(-1+4^s).
\end{equation}
Notice that the right side of the above inequality approaches 0 when both $s
\rightarrow \infty$ and $s \rightarrow 0$, thus we can conclude that the
numerator of \eqref{vol2qbms} will be $< 1$ and therefore, the value
of $\omega$ (be it either $> 1$ or $< 1$) will be completely dependent on
the explicit choice of the value of $s$.  

The important point to recognize here is not the symplex calculations
but rather the flag manifold volume.  Through the Euler
parameterization of $SU(N)$ and $U(N)$ 
given in \cite{Tilma2,UandCPN}
%given in Chapters \ref{chap:chapter-suN} and \ref{chap:UandCPN}
we have been able to generate
the appropriate representation of the ``truncated'' Haar measure which
is crucial to any mixed state volume calculation.  It is this factor
which is not ``user dependent''; i.\ e.\ dependent on the initial
distribution chosen for the ($N-1$)-dimensional symplex, and therefore
not completely subject to disagreements between researchers studying
entanglement.\footnote{Although disagreements in numerical
values are found, they are mostly due to variations in the ranges of the 
$N(N-1)$ parameters which define the measure (see for example
\cite{SlaterNew} and references within).}

%--------------------------------------------------------------------

%--------------------------------------------------------------------
\section{Qubit/Qutrit Entanglement}
%---
%---
\subsection{Pure State Entanglement}

By following the same procedure as was done in the two
qubit case, we can derive the manifold of operations that produce
entanglement of an initial qubit/qutrit pure state
\begin{equation}
\rho_d=
\begin{pmatrix}
1 & 0 & 0 & 0 & 0 & 0\\
0 & 0 & 0 & 0 & 0 & 0\\
0 & 0 & 0 & 0 & 0 & 0\\
0 & 0 & 0 & 0 & 0 & 0\\
0 & 0 & 0 & 0 & 0 & 0\\
0 & 0 & 0 & 0 & 0 & 0
\end{pmatrix}
\end{equation}
via a $U \in SU(6)$ given in
%Chapter \ref{chap:suN}
\cite{Tilma2}
\begin{align}
U =&\;
U(\alpha_1,\alpha_2,\alpha_3,\alpha_4,\alpha_5,\alpha_6,\alpha_7,\alpha_8,\alpha_9,\alpha_{10},\alpha_{11},0,
\alpha_{13},0,\alpha_{15},0,\alpha_{17}, \nonumber \\
&\times 0,\alpha_{19},0,\alpha_{21},0,\alpha_{23},0,\alpha_{25},0,\alpha_{27},
0,\alpha_{29},0,\alpha_{31},\alpha_{32},\alpha_{33},\alpha_{34},\alpha_{35}) \nonumber \\
U=&\;e^{i\lambda_{3}\alpha_{1}}e^{i\lambda_{2}\alpha_{2}}
 e^{i\lambda_{3}\alpha_{3}}e^{i\lambda_{5}\alpha_{4}} 
 e^{i\lambda_{3}\alpha_{5}}e^{i\lambda_{10}\alpha_{6}}
 e^{i\lambda_{3}\alpha_{7}}e^{i\lambda_{17}\alpha_{8}}
 e^{i\lambda_{3}\alpha_{9}}e^{i\lambda_{26}\alpha_{10}}
\nonumber \\
 &\times e^{i\lambda_{3}\alpha_{11}} 
 e^{i\lambda_{3}\alpha_{13}} 
 e^{i\lambda_{3}\alpha_{15}}
 e^{i\lambda_{3}\alpha_{17}} 
 e^{i\lambda_{3}\alpha_{19}}
 e^{i\lambda_{3}\alpha_{21}}
 e^{i\lambda_{3}\alpha_{23}} \nonumber \\
&\times e^{i\lambda_{3}\alpha_{25}} 
 e^{i\lambda_{3}\alpha_{27}}
 e^{i\lambda_{3}\alpha_{29}}
 e^{i\lambda_{3}\alpha_{31}}
 e^{i\lambda_{15}\alpha_{33}}e^{i\lambda_{24}\alpha_{34}} 
 e^{i\lambda_{35}\alpha_{35}},
\end{align}
by taking the partial transpose of $U\rho_d U^\dagger$ and evaluating the
corresponding eigenvalues through the Peres-Horodecki
criterion.\footnote{We are forced to generate the six eigenvalue equations rather than evaluate
the constant term from the characteristic polynomial because in
this case we have two eigenvalues equal to zero thus negating the
constant term's effectiveness.} 
Doing this work, which did not yield well to simplification, generates
the following \textit{hypothesized} manifold
\begin{equation}
\label{psentanglingmanifoldsu6}
\EuScript{M}_{\text{E}_{SU(6)}} = \frac{SU(6)}{U(5)\times
U(1)_{SU(2)_{\kappa}}} = \frac{\mathbb{C}\mbox{P}^5}{U(1)_{SU(2)_{\kappa}}},
\end{equation}
where $\kappa\cong -2(2\alpha_1+\alpha_3+\alpha_7+\alpha_9)$ is the cumulative 
phase.\footnote{After taking into account the degeneracy in the eigenvalues, full simplification of the resulting 4th order
characteristic polynomial was not possible without making certain numerical assumptions.  Thus, the
actual representation of $\kappa$ as a function of $\alpha_1, \alpha_3, \alpha_7$ and $\alpha_9$ 
was not possible; thus the equivalence.} 
The volume of the manifold is 
\begin{align}
V_{\EuScript{M}_{\text{E}_{SU(6)}}} =V\biggr(\frac{SU(6)}{U(5)\times U(1)_{SU(2)_{\kappa}}}\biggl)
&= \frac{V_{\mathbb{C}\text{P}^5}}{10*V_{U(1)_{\lambda_3}}} \nonumber \\
&= \frac{\pi^5}{5! *20\pi^4} \nonumber \\
&= \frac{\pi}{2400}.
\label{eq:psentanglingmanifoldvolsu6}
\end{align}
If we make the same argument for the volume of this manifold to be equivalent to the
volume of entangled qubit/qutrit pure states, as we did in the two qubit case,
then, in this case, our values are within the ranges specified in
\cite{ZyczkowskiV1,ZyczkowskiV2} but since we do not include the
possibility of bound entangled states (states which are entangled, but
which have positive partial transposes) we must concede that our value
given in equation (\ref{eq:psentanglingmanifoldvolsu6}) is probably
too small.
%---
%---
\subsection{Mixed State Entanglement}

Again, as in the two qubit case,
one can see that for a general $\rho_d$ for a qubit/qutrit system 
%defined in Chapter \ref{chap:suN}
\cite{Tilma2}
\begin{align}
\rho_d = \text{diag}\{&{\sin({{\theta}_1})}^2\,{\sin({{\theta
}_2})}^2\,{\sin({{\theta}_3})}^2\,{\sin({{\theta}_4})}^2\,{\sin({{\theta}_5})}^2,
{\cos({{\theta}_1})}^2\,{\sin({{\theta}_2})}^2\,{\sin({{\theta}_3})}^2\,{\sin({{\theta}_4})}^2\,{\sin({{\theta}_5})}^2,
\nonumber \\
&{\cos({{\theta}_2})}^2\,{\sin({{\theta}_3})}^2\,{\sin({{\theta}_4})}^2\,{\sin({{\theta}_5})}^2,
{\cos({{\theta}_3})}^2\,{\sin({{\theta}_4})}^2\,{\sin({{\theta}_5})}^2,{\cos({{\theta}_4})}^2\,{\sin({{\theta}_5})}^2,
{\cos({{\theta}_5})}^2\}
\end{align}
when $\theta_i \neq \pi/2$, ($i=1,\ldots, 5$)
the generalization of equation \eqref{wooterscond} for the
qubit/qutrit case \cite{ZyczkowskiV1,ZyczkowskiV2}
\begin{equation}
\label{wooterscondsu6}
Tr[\rho^2] > \frac{1}{5}
\end{equation}
where
\begin{align}
Tr[\rho^2]=&Tr[(U\rho_d U^\dagger)^2]=Tr[\rho_d^2] \nonumber \\
=&\,1-2{\sin({{\theta}_5})}^2\,
(1+{\sin({{\theta}_5})}^2-{\sin({{\theta}_4})}^2\,{\sin({{\theta}_5})}^2+
{\sin({{\theta}_4})}^4\,{\sin({{\theta}_5})}^2 \nonumber \\
&-{\sin({{\theta}_3})}^2\,{\sin({{\theta}_4})}^4\,{\sin({{\theta}_5})}^2+
{\sin({{\theta}_3})}^4\,{\sin({{\theta}_4})}^4\,{\sin({{\theta}_5})}^2-
{\sin({{\theta}_2})}^2\,{\sin({{\theta}_3})}^4\,{\sin({{\theta}_4})}^4\,
{\sin({{\theta}_5})}^2 \nonumber \\
&+{\sin({{\theta}_2})}^4\,{\sin({{\theta}_3})}^4\,
{\sin({{\theta}_4})}^4\,{\sin({{\theta}_5})}^2-
{\sin({{\theta}_1})}^2\,{\sin({{\theta}_2})}^4\,{\sin({{\theta}_3})}^4\,
{\sin({{\theta}_4})}^4\,{\sin({{\theta}_5})}^2\nonumber \\
&+{\sin({{\theta}_1})}^4\,{\sin({{\theta}_2})}^4\,{\sin({{\theta}_3})}^4\,
{\sin({{\theta}_4})}^4\,{\sin({{\theta}_5})}^2)
\end{align}
could still be satisfied.  In this case then
we would have for the entangled mixed state product measure (under
appropriate ranges for $\Lambda_i$ and $\alpha_i$)
\begin{align}
dV_{E_{ms}}=&\alpha_s
\Lambda_1^{s-1}\cdots\Lambda_5^{s-1}(1-\sum_{i=1}^5\Lambda_i)^{s-1}d\Lambda_1\ldots
d\Lambda_5 \nonumber \\
&\times d\biggr(\frac{SU(6)}{U(1)_{SU(2)} \times U(1)_{SU(3)} \times 
U(1)_{SU(4)} \times U(1)_{SU(5)} \times U(1)_{SU(6)}}\biggl)
d\alpha_{30} \ldots d\alpha_1.
\end{align}
Another general (and again, rather naive) evaluation of this measure for our situation yields
\begin{align}
\label{volqbqtms}
V_{E_{ms}}&= \alpha_s\,\frac{\left( {{a_L}}^s - {{a_U}}^s \right) \,
    \left( {{b_L}}^s - {{b_U}}^s \right) \,
    \left( {{c_L}}^s - {{c_U}}^s \right) \,
    \left( {{d_L}}^s - {{d_U}}^s \right) \,
    \left( {{e_L}}^s - {{e_U}}^s \right) \,
    \left( {{f_L}}^s - {{f_U}}^s \right) }{s^6} \times
    \frac{\pi^{15}}{34560} \nonumber \\
&= \omega\,\frac{\pi^{15}}{34560},
\end{align}
where $\{a_U,a_L\}$ through $\{f_U,f_L\}$ are the squared values of
the maximal and minimal ranges of $\Lambda_i$ that satisfy
\eqref{wooterscondsu6}.  As before,
the symplex measure must be \textit{assumed} to be non-zero, therefore using
the work contained in \cite{UandCPN}
%equation (\ref{eq:2qbomega}) from Chapter \ref{chap:UandCPN}
we can \textit{hypothesize} that $\omega$ has the following bounds (dependent on the value of $s$ and
recalling that $\alpha_s > 0$ and $s >0$)
\begin{equation}
0 <  \left( {{a_L}}^s - {{a_U}}^s \right) \,
    \left( {{b_L}}^s - {{b_U}}^s \right) \,
    \left( {{c_L}}^s - {{c_U}}^s \right) \,
    \left( {{d_L}}^s - {{d_U}}^s \right) \,
    \left( {{e_L}}^s - {{e_U}}^s \right) \,
    \left( {{f_L}}^s - {{f_U}}^s \right) < 6^{-6s}(-1+6^s).
\end{equation}
Again we notice that the right side of the above inequality approaches 0 when both $s
\rightarrow \infty$ and $s \rightarrow 0$, we can again conclude that the
numerator of \eqref{volqbqtms} will be $< 1$ and therefore, the value
of $\omega$ (be it either $> 1$ or $< 1$) will be completely dependent on
the explicit choice of the value of $s$.  
Also, as before in the two qubit case, the important point to recognize here is not the symplex calculations
but rather the flag manifold volume. 

%--------------------------------------------------------------------

%--------------------------------------------------------------------
\section{Conclusions}

In this paper we have applied our $SU(N)$ and $U(N)$
parameterizations to the  
two qubit and qubit/qutrit system in order to explicit calculate the 
manifold of operations which entangle two qubit and qubit/qutrit pure
states.  We have also been able to give the volume of this manifold, as
well as the \textit{hypothesized} volume for the set of all entangled
two qubit and qubit/qutrit pure and mixed states.  In the pure state
case, the values were within the ranges given by
\cite{ZyczkowskiV1,ZyczkowskiV2} but in the qubit/qutrit case, because
we did not take into account the possibility of bound entangled states
(which do not appear in the two qubit case) our volume is \textit{most likely}
smaller than the actual volume for the set of all entangled
qubit/qutrit pure states.  

Work is continuing on the mixed state situation; explicitly in
calculating the volume of the mixed state manifold without having to
know the exact probability distribution on the ($N-1$)-dimensional
symplex.  Extensions of the pure state work to two qutrit systems is
also ongoing.

%--------------------------------------------------------------------

%--------------------------------------------------------------------
\section*{Acknowledgments}

We would like to thank Dr.\ M.\ Byrd for his editorial help on 
the various representational conventions for density matrices as
well as Anil Shaji for invaluable assistance in calculating the various
manifold volumes given here.

%--------------------------------------------------------------------
%--------------------------------------------------------------------
\appendix
%--------------------------------------------------------------------
%--------------------------------------------------------------------
\section{$SU(4)$ Lie Algebra}
\label{app:su4lambdas}

From \cite{Tilma1} we know that the
Gell-Mann type basis for the Lie algebra of
$SU(4)$ is given by the following set of matrices \cite{Greiner}:
\begin{equation}
\begin{array}{crcr}
\lambda_1 = \left( \begin{array}{cccc}
                     0 & 1 & 0 & 0 \\
                     1 & 0 & 0 & 0 \\
                     0 & 0 & 0 & 0 \\
                     0 & 0 & 0 & 0  \end{array} \right), &
\lambda_2 = \left( \begin{array}{crcr} 
                     0 & -i & 0 & 0 \\
                     i &  0 & 0 & 0 \\
                     0 &  0 & 0 & 0 \\
                     0 &  0 & 0 & 0  \end{array} \right), &
\lambda_3 = \left( \begin{array}{crcr} 
                     1 &  0 & 0 & 0 \\
                     0 & -1 & 0 & 0 \\
                     0 &  0 & 0 & 0 \\
                     0 &  0 & 0 & 0  \end{array} \right), \\
\lambda_4 = \left( \begin{array}{clcr} 
                     0 & 0 & 1 & 0 \\
                     0 & 0 & 0 & 0 \\
                     1 & 0 & 0 & 0 \\
                     0 & 0 & 0 & 0  \end{array} \right), &
\lambda_5 = \left( \begin{array}{crcr} 
                     0 & 0 & -i & 0 \\
                     0 & 0 &  0 & 0 \\
                     i & 0 &  0 & 0 \\
                     0 & 0 &  0 & 0 \end{array} \right), &
\lambda_6 = \left( \begin{array}{crcr} 
                     0 & 0 & 0 & 0 \\
                     0 & 0 & 1 & 0 \\
                     0 & 1 & 0 & 0 \\
                     0 & 0 & 0 & 0 \end{array} \right), \\
\lambda_7 = \left( \begin{array}{crcr} 
                     0 & 0 &  0 & 0 \\
                     0 & 0 & -i & 0 \\
                     0 & i &  0 & 0 \\
                     0 & 0 &  0 & 0 \end{array} \right), &
\lambda_8 = \frac{1}{\sqrt{3}}\left( \begin{array}{crcr} 
                     1 & 0 &  0 & 0 \\
                     0 & 1 &  0 & 0 \\
                     0 & 0 & -2 & 0 \\
                     0 & 0 &  0 & 0  \end{array} \right), &
\lambda_9 = \left( \begin{array}{crcr} 
                     0 & 0 & 0 & 1 \\
                     0 & 0 & 0 & 0 \\
                     0 & 0 & 0 & 0 \\
                     1 & 0 & 0 & 0 \end{array} \right), \\
\lambda_{10} = \left( \begin{array}{crcr} 
                     0 & 0 & 0 & -i \\
                     0 & 0 & 0 &  0 \\
                     0 & 0 & 0 &  0 \\
                     i & 0 & 0 &  0  \end{array} \right), &
\lambda_{11} = \left( \begin{array}{crcr} 
                     0 & 0 & 0 & 0 \\
                     0 & 0 & 0 & 1 \\
                     0 & 0 & 0 & 0 \\
                     0 & 1 & 0 & 0\end{array} \right), &
\lambda_{12} = \left( \begin{array}{crcr} 
                     0 & 0 & 0 &  0 \\
                     0 & 0 & 0 & -i \\
                     0 & 0 & 0 &  0 \\
                     0 & i & 0 &  0  \end{array} \right), \\
\lambda_{13} = \left( \begin{array}{crcr} 
                     0 & 0 & 0 & 0 \\
                     0 & 0 & 0 & 0 \\
                     0 & 0 & 0 & 1 \\
                     0 & 0 & 1 & 0  \end{array} \right), &
\lambda_{14} = \left( \begin{array}{crcr} 
                     0 & 0 & 0 &  0 \\
                     0 & 0 & 0 &  0 \\
                     0 & 0 & 0 & -i \\
                     0 & 0 & i &  0  \end{array} \right), &
\lambda_{15} = \frac{1}{\sqrt{6}}\left( \begin{array}{crcr}
                     1 & 0 & 0 &  0 \\
                     0 & 1 & 0 &  0 \\
                     0 & 0 & 1 &  0 \\
                     0 & 0 & 0 & -3  \end{array} \right).
\end{array}    
\end{equation}
Using these matrices one can then generate the various group operations given in 
section \ref{sec:entanglingops}.  Similarly, in \cite{Tilma2} one can see
how to construct the $N^2-1$ elements of the $SU(N)$ Lie algebra necessary for general
$SU(N)$ group operations.
%--------------------------------------------------------------------

%--------------------------------------------------------------------

\bibliographystyle{apsrev}
\bibliography{main}

%--------------------------------------------------------------------
\end{document}